\documentclass{aa}  
\usepackage{graphicx}
\usepackage{epstopdf}
\usepackage{txfonts}
\usepackage{xcolor}
\usepackage{ulem}

\usepackage[nameinlink,capitalise]{cleveref}
\crefname{section}{Sect.}{Sects.}
\Crefname{section}{Section}{Sections}
\crefname{figure}{Fig.}{Figs.}
\Crefname{figure}{Figure}{Figures}
\crefname{equation}{Eq.}{Eqs.}
\Crefname{equation}{Equation}{Equations}

\def\tein{\theta_{\mathrm{E}}}
\def\NSL{\mathcal{N}_{\rm{SL}}}
\def\asps{\alpha_{\rm{SPS}}}
\def\gammadm{\gamma_{\rm{DM}}}
\def\muE{\mu_{\tau_E}}
\def\sigE{\sigma_{\tau_E}}

\def\pr{{\rm P}}

\def\Nlensfull{12,418}
\def\Nlens{124}

\newcommand*{\Euclid}{\textit{Euclid}\xspace}

\begin{document}

   \title{Constraining galaxy properties with complete samples of lenses}
   

   \author{Qing Zhou
          \inst{\ref{leiden}}
          \and
          Alessandro Sonnenfeld\inst{\ref{sjtu1},\ref{sjtu2},\ref{sjtu3},\ref{leiden}}\fnmsep\thanks{\email{sonnenfeld@sjtu.edu.cn} }\and
          Henk Hoekstra\inst{\ref{leiden}}
          }

   \institute{Leiden Observatory, Leiden University, P.O. Box 9513, 2300 RA Leiden, The Netherlands\label{leiden} \and
    Department of Astronomy, School of Physics and Astronomy, Shanghai Jiao Tong University, Shanghai 200240, China\label{sjtu1} \and
Shanghai Key Laboratory for Particle Physics and Cosmology, Shanghai Jiao Tong University, Shanghai 200240, China\label{sjtu2} \and
Key Laboratory for Particle Physics, Astrophysics and Cosmology, Ministry of Education, Shanghai Jiao Tong University, Shanghai 200240, China\label{sjtu3}
             }

   \date{Received ; accepted }

 
  \abstract{
The statistics of Einstein radii for a sample of strong lenses can provide valuable constraints on the underlying mass distribution. The correct interpretation, however, relies critically on the modelling of the selection of the sample, which has proven to be a limiting factor. This may change thanks to upcoming uniform high-resolution imaging surveys that cover a large fraction of the sky, because they can provide complete lens samples, with well understood selection criteria.
To explore how the observed distribution of Einstein radii depends on the galaxy properties, we simulated a realistic complete sample of strong lenses, predicting a number density of lenses of about 2.5 deg$^{-2}$ for a \Euclid-like setup. Such data can break the degeneracy between the stellar initial mass function (IMF) and the inner slope of the density profile of dark matter, without having to rely on additional information from stellar dynamics. 
We found that a survey covering only 50 deg$^2$ can already provide tight constraints: assuming that the cosmology is known, the dark matter slope is recovered with an uncertainty of $3.5\%$, while the uncertainty in the ratio between the true stellar mass and that inferred from stellar population modelling was found to be $10\%$. These findings highlight the potential of this method when applied to samples of lenses with well-understood selection functions.}


   \maketitle
%
\section{Introduction}

Understanding how galaxies populate dark matter halos is one of the most fundamental questions in the field of astrophysics and cosmology, as it is crucial for testing cosmological models, understanding the growth of cosmic structures, and probing the nature of dark matter and dark energy. A key element in models of galaxy formation is the relation between the mass in stars and the dark matter mass. A robust determination of this relation remains a challenge for a number of reasons. First, the accuracy of stellar mass estimates is currently limited. The most important source is our limited knowledge of the stellar initial mass function (IMF): the inferred stellar mass can vary by as much as a factor of two for different IMF choices. Moreover, there is no consensus on whether the IMF is universal or varies as a function of galaxy type \citep[see][and references therein]{Smi20}. More generally, the IMF is important for our understanding of the baryon cycle, as it affects the inferred star formation history and the associated feedback processes. 

Uncertainties in the distribution of dark matter also contribute to the problem. Apart from fundamental questions about the nature of dark matter, the physical processes affecting galaxy evolution also influence the distribution of dark matter within halos. On the scales of massive galaxies, hydrodynamical simulations that include cold dark matter generally predict cuspy inner density profiles \citep[see e.g.,][]{2010duffy, 2015schaller, Xu++17, Pei++17}, with a density slope that is sensitive to the relative importance between gas infall and feedback \citep{2014dicintio}. In principle, the inner distribution of dark matter can be obtained by means of dynamical modelling of kinematic data \citep{2012cappellari}, although in practice such measurements are currently only feasible in the nearby Universe.

Strong gravitational lensing is one of the few probes that allows direct mass measurement of massive galaxies at cosmological distances. However, the presence of degeneracies in mass modelling and the intrinsic uncertainty associated with source properties are among the obstacles that must be overcome. To break degeneracies in the lens modelling, additional constraints from stellar kinematics have been included \citep[e.g.,][]{2002treu&koopmans,2003koopmans&treu,2004treu&koopmans,auger2010a,auger2010b,2015sonnenfeld,Shajib2021}, but to simultaneously fit lensing and kinematic data, more complex models are required: dynamical analyses need to take into account the orbital anisotropy of the tracers and the 3-dimensional distribution of mass. If not properly modelled, these features can be sources of systematic errors.

An alternative approach is to use a statistical sample of strong lenses, which can provide constraints for an ensemble of galaxies, while avoiding the need for additional observables \citep{2021sonnenfeld_paper1}. In this case, the constraining power depends critically on the information that is available for the individual lenses. For instance, \citet{2021sonnenfeld_paper1} showed that the degeneracy between stellar and dark matter mass can be broken if the positions and the magnification ratios between the multiple images can be measured. However, it remains to be seen whether magnification ratios can be measured to within a few percent for large samples of lenses, because accurate knowledge of this information requires detailed modelling of the images of a lensed extended source \citep{Suyu2012,2018sonnen}. Without magnification information it is very difficult to make accurate inferences. 

In this paper we explore the prospects of using another piece of information, namely the total number of lenses in a given survey.
To date, the observed number of lenses has predominantly been used as a way to constrain cosmology \citep[e.g.,][]{Koc96,Wam++04,Ogu++08,boldrin2016,Har20}, under the assumption that the properties of the mass distribution of the lenses are known. However, as uncertainties on the cosmological parameters become smaller, the opposite approach becomes possible: at fixed cosmology, the number of lenses can be used to infer the galaxy properties. The main reason why such a strategy has not yet been adopted is that the lens samples from past surveys are incomplete, with ill-defined selection functions. 

Ongoing and upcoming surveys such as \Euclid \citep{Mellier2024}, the Rubin Observatory Legacy Survey of Space and Time (LSST; \citealt{2020vera_rubin}), the Nancy Grace Roman Space Telescope \citep[e.g.,][]{2021MNRAS.507.1746E,2021arXiv211103081R} and the Chinese Space Station Telescope (CSST; e.g., \citealt{Zhancsst2011,Caocsst2018,csst_zhanhu21}) will cover vast areas of the extragalactic sky. Thanks to the unprecedented image quality,  especially from space, these projects will increase the number of galaxy-galaxy strong lenses by at least two orders of magnitude \citep{2014serjeant,collett2015,Cao2023}. With such statistics, it might be possible to identify regions in the parameter space of observable lens properties for which a lens survey is complete, meaning that every lens that satisfies a specified cut in observational space is included in the survey.
The main advantage of working with a complete sample is that the selection function can be easily modelled.\footnote{If the sample is complete, the probability of finding a detectable lens is one.}
As a result, strong lensing selection effects can be robustly accounted for \citep{2022sonnenfeld_paper3}, making it easier to interpret observable quantities, such as the total number of lenses and their distribution of Einstein radii.\footnote{For axisymmetric lenses, the Einstein radius, $r_{\rm E}$, is the radius within which the mean surface density corresponds to the critical surface mass density $\Sigma_{\rm{crit}}$.} 

In this work, we explore the potential of a complete sample of strong lenses to constrain the stellar IMF and the inner slope of the dark matter density profile for an ensemble of elliptical galaxies. In \cref{sec:lens_sim} we describe how we simulated realistic samples of lenses and sources, and analyse the properties of the simulated lenses. In \cref{sec:results} we show that it is possible to break the IMF-dark matter degeneracy from the distribution of Einstein radii of the lens samples, without additional constraints from stellar dynamics or magnification information. In our analysis, we assume that the uncertainties in cosmological parameters have a negligible effect on the number of strong lenses. In \cref{sec:cosmology} we explore the validity of this assumption. In \cref{sec:discussion} we discuss the implications of our findings and present our conclusions.
The Python code used in this work is made publicly accessible here.\footnote{\url{https://github.com/fatginger1024/SLens}}

\section{Simulations} 
\label{sec:lens_sim}

To explore how well a complete sample of strong lenses can constrain the IMF and inner slope of the dark matter halo, we generated a catalogue of synthetic lenses and compared the lensing statistics to predicted distributions for which we varied the properties of their inner structure. In this section we describe how we generated a realistic sample of lenses for a setup that resembles \Euclid in terms of imaging depth, and analyse the galaxy properties of our mocks. Throughout this paper, we assume that all lenses that meet our criteria are detected.

\subsection{Lens mass properties}
\label{sec:sims}

The statistics of Einstein radii depends on the distribution of the stellar and dark matter in the lenses. To create realistic mock samples, we use the Marenostrum Institut de Ci{\`e}nces de l'Espai Grand Challenge (MICE-GC) simulation \citep{Fosalba2015a,Crocce2015, Fosalba2015b}, which provides realistic galaxy and halo properties up to $z=1.4$ over an area of 5000 deg$^2$. The large area allows us to subdivide the sample into 100 subsets of 50 deg$^2$, enabling a robust statistical analysis while also capturing field-to-field variations. 

The mock galaxy catalogue is derived from the MICE-GC N-body simulation, which contains about 70 billion dark matter particles in a comoving volume of approximately $(3 h^{-1}\mathrm{Gpc})^3$ \citep{Fosalba2015a}. Thanks to the particle mass of $m_{\rm p}=2.93\times 10^{10} h^{-1}M_\odot$ it is possible to identify halos with masses $M>3\times 10^{12} h^{-1}M_\odot$, sufficient to construct samples of strong lenses. As described in \cite{Carretero2015} and \cite{Crocce2015}, the galaxy properties were assigned to the dark matter halos using a hybrid method, combining a halo occupation distribution (HOD; e.g., \citealt{1998jing&mo,2000peacock&smith,2002berlind&weinberg}) with sub-halo abundance matching (SHAM; e.g., \citealt{2004vale&ostriker,2006conroy&wechsler,2010guo&white}). Additional observational constraints were imposed on the luminosity function and colour distribution in order to match data from the Sloan Digital Sky Survey (SDSS; \citealt{2000sdss}). 
The resulting  galaxy catalogue also reproduces the observed galaxy clustering properties as a function of luminosity and colour.

Throughout this work, we only consider the lensing effect of central galaxies and their dark matter halos, while ignoring the contribution from satellites. This is a good approximation for galaxy-scale lenses, which dominate the lens population. Group and cluster-scale lenses often present multiple galaxies within the Einstein radius, but they constitute a minority of lenses. In a real-world application, they can be easily identified and removed from both the observed and simulated samples.
 
To keep the lens statistics analyses within reach of analytic forms, we assume that all foreground lenses are axisymmetric, consisting of a stellar component and a dark matter component. We expect this simplification to have a minimal impact on our results, because the Einstein radius and the lensing cross-section of a lens are a very weak function of lens ellipticity \citep[see section 3.3 of][]{Son++23}. 
We adopted a two-component mass model for our lenses, consisting of a stellar component and a dark matter halo. For the dark matter distribution we used a generalised NFW (gNFW) profile, which allows for the parametrisation of the mass distribution in the inner region with three degrees of freedom \citep[e.g.,][]{zhao1996}. The density profile of a gNFW profile takes the following parametric form
\begin{equation}
\rho(r)  = \frac{\rho_{\rm s}}{(r/r_{\rm s})^{\gamma_{\rm{DM}}}(1+r/r_{\rm s})^{3-\gamma_{\rm{DM}}}}, \label{density gnfw}
\end{equation}
with $\rho_{\rm s}\equiv \delta_c\rho_{\rm{crit}}$, where 
$\rho_{\rm{crit}}$ is the critical density of the universe.\footnote{The critical density of the universe is defined as $\rho_{\rm{crit}} = 3H^2(z)/(8\pi G)$ at the redshift $z$ of interest.} The inner region falls off as $\rho(r)\sim r^{-\gamma_{\rm{DM}}}$ and the outer region as $\rho(r)\sim r^{-3}$, and the scale radius $r_{\rm s}$ determines where the transition occurs. The characteristic overdensity, $\delta_c$, is a function of the concentration parameter $c$ 
\begin{equation}
\delta_c = \frac{200}{3}\frac{c^{\gamma_{\rm{DM}}}(3-\gamma_{\rm{DM}})}{{}_2F_1(3-\gamma_{\rm{DM}},3-\gamma_{\rm{DM}},4-\gamma_{\rm{DM}};-c)},
\end{equation}
where ${}_2F_1 (a, b; c; x)$ is the hypergeometric function \citep[e.g.,][]{keeton2001}, and $c$ is defined as 
 \begin{equation}
    c = \frac{r_{200}}{r_{\rm s}},
 \end{equation}
where $r_{200}$, is the virial radius of the dark matter halo.\footnote{The virial radius $r_{200}$ is the radius enclosing the region within which the mean density is 200 times the critical density. The corresponding enclosed mass is $M_{200}$.}

Simulations have shown that the average concentration is a relatively weak function of mass, but where the normalisation depends on the cosmology.
To determine the halo concentrations, we adopted the mass-concentration-redshift (MCR) relation derived by \cite{ludlowa2016}
for the {\it Planck} cosmology from \cite{2015planck}.
For a given virial mass, the values for $c$ follow a log-normal distribution, for which we took a constant value of $0.088$ \citep[see Fig. 10 of][for details]{ludlowa2016}. Hence, the concentration parameter of a galaxy is assigned based on its halo mass and redshift. Consequently, the dark matter density profile is described by three parameters: the halo mass $M_{200}$, the concentration parameter $c$ and the dark matter inner slope $\gammadm$. 

We used the same procedure to obtain halo masses and concentrations for both the observed and the model lens populations. This is equivalent to assuming that the stellar-to-halo mass relation and mass-concentration relations are known exactly during the modelling stage. 
The rationale for this assumption, which enables us to keep the dimensionality of the problem low, is that these properties of the dark matter distribution can be obtained from complementary observations, such as weak lensing and galaxy clustering. In \cref{sec:discussion} we discuss the impact of this assumption and possible ways to avoid it.

For the stellar component of our lens model, we adopted a S\'{e}rsic profile. Its projected stellar mass distribution is given by 
\begin{equation}
\Sigma(R) = \Sigma_0 \exp\left[-b(n)\,(R/R_{\rm{e}})^{1/n}\right],
\end{equation}
with 
\begin{equation}
\Sigma_0 = \frac{M_*b(n)^8}{8\pi R_{\rm{e}}^2\Gamma(8)},
\end{equation}
where $b(n)= 2n-\frac{1}{3}+\frac{4}{405n}+\frac{46}{25515n^2}+\mathcal{O}(n^3)$, and $R_{\rm{e}}$ is the half-light radius. We assigned a S\'{e}rsic index of $n=4$ for all of our lens galaxies, which is a representative value for bright elliptical galaxies \citep{ciotti1999Relaw}. 

To describe the stellar mass of a galaxy we make the following distinctions. We indicate with $M_*^{\rm true}$ the true stellar mass of a galaxy, and with $M_*^{\rm{obs}}$ the observed stellar mass, obtained by fitting a stellar population synthesis (SPS) model to the photometric data. The former is needed in the lens modelling process. The MICE catalogue, however, is based on the latter. Therefore we define an IMF mismatch parameter, $\alpha_{\rm SPS}$, as the ratio of the true stellar mass to the observed stellar mass as determined by an SPS model 
\begin{equation}
\alpha_{\rm{SPS}} \equiv \frac{M_*^{\rm{true}}}{M_*^{\rm{obs}}}.
\end{equation}

Given the stellar mass from the MICE catalogue and a value for $\asps$, we can assign a true stellar mass to each galaxy. Another important parameter for lens modelling is the half-light radius. In this work we assigned the half-light radii based on the mass-size relation for massive ellipticals from \cite{2019bsonnenfeld}. In summary, the stellar mass profile is parametrised by the lens galaxy redshift $z$, the observed stellar mass $M_*^{\rm{obs}}$, the half-light radius $R_{\rm{e}}$ and the stellar mass mismatch parameter $\asps$.  

We generated the foreground galaxies from the central galaxies of MICE with observed stellar mass $M_*^{\rm{obs}
}>10^{11}\,h^{-1}\,\rm{M_{\odot}}$ at $z<1$. We assigned a universal value to the dark matter inner slope and the IMF mismatch parameter to generate the mock observed sample with $\gamma_{\rm{DM}}^{\rm{true}}=1.3$ and $\alpha_{\rm{DM}}^{\rm{true}}=1.2$. We assumed a flat $\Lambda \rm{CDM}$ cosmology with: $\Omega_{\rm m}=0.25$, $\Omega_{\Lambda}=0.75$, and a present-day Hubble constant $H_0=73\,\rm{km\,s^{-1}\,Mpc^{-1}}$.  

\subsection{Background source population}

The strong lensing statistics also depend on the properties of the background sources that can be detected. Hence, the precise numbers will depend on the adopted survey characteristics. Here, we consider a setup that can mimic the performance for space-based surveys such as \Euclid and CSST. To this end, we generated background sources using the COSMOS2015 catalogue \citep{cosmos2016}, which contains precise photometric redshifts ($\sigma({\Delta z/(1+z)})=0.007$)  and stellar mass measurements for more than half a million objects over 2~deg$^2$, with a limiting $r$-band magnitude of $26.5$ ($3\sigma$ for a $3^{\prime\prime}$ diameter aperture). For sources at $z_{\rm s}>1$, this photometric redshift uncertainty translates into an error in Einstein radius of less than one percent, which has a negligible effect on the generated lens samples. Moreover, the depth and redshift coverage are sufficient to simulate the majority of sources that would be detected by \Euclid and CSST \citep{collett2015,Cao2023}. We approximated the background sources as point-like: this is appropriate, since typical strongly lensed sources have much smaller sizes compared to the Einstein radius of their lenses \citep[see e.g.][]{2019bsonnenfeld}. In \cref{sec:discussion} we discuss how this can be avoided in a real survey.

\subsection{Strong lensing events}

Our aim is to explore the constraining power of a strong lensing survey based on data from a space-based mission such as \Euclid or the CSST. This enables us to define the main characteristics of a strong lensing system that would be detected in such observations.
Given a simulated lens-source pair, we defined a strong lens as a system for which:

\begin{itemize}
    \item At least two images of the same source are produced;
    \item The observed apparent magnitudes of the images are brighter than a limiting magnitude $m_{s}=25$;
    \item The Einstein radius of the lens is larger than $0\farcs5$.
\end{itemize}

The rationale for the third criterion is that strong lenses can more easily be identified if the multiple images are well resolved. Tests on strong lens image simulations show that the Einstein radius is the main property that determines the probability of a lens with detected arcs to be discovered by a lens-finding algorithm \citet{Gav++14, HOV23}. Therefore we set a lower limit to the Einstein radius of the sample, selecting only lenses with $\theta_E\geq 0\farcs5$. This value of the minimum Einstein radius is slightly larger than that of the smallest image separation galaxy-galaxy lenses detected in space-based imaging observations \citep{2022garvin}. For the \textit{Euclid} wide survey, the 5$\sigma$ detection limit in the VIS band is about $m_{\rm AB}=26.2$ \citep{2022scaramella}. 
We set the limiting magnitude $m_{s}$ to be well above this value, to simulate a scenario in which only the most obvious lenses are used in the analysis.
The implications of the criteria, as well as their applicability to real samples of lenses will be further discussed in \cref{sec:discussion}.

To generate the mock observed sample, for each foreground galaxy
we first computed its maximum cross-section area, defined as the area on the source plane enclosed by the radial caustic at the limiting source redshift $z=6$. Then, we placed sources randomly in a cone with base area equal to the maximum cross-section area and solved the lens equation for each source within the cone. Finally, we applied the criteria introduced above to determine whether each galaxy is a lens or not.
Our fiducial analysis was carried out with a sample of lenses complete down to a minimum Einstein radius of $0\farcs 5$. In addition, we considered a more conservative sample, complete only down to $\tein=1''$.

One aspect that we did not consider in our lens definition is the fact that, in real surveys, images that lie close to the centre of the lens are more difficult to detect, due to the contamination from the lens light. While this is an important issue for ground-based searches, we believe it to be a minor effect for a space-based survey, especially when working in the more conservative limit of $\tein>1''$. This is because we only considered lensed images that are significantly brighter than the detection limit, meaning that they would be easily detected even in the presence of significant lens light contribution.
However, for a survey like the LSST, the smearing by the point spread function significantly reduces the observed surface brightness of lensed images, in turn reducing the contrast between source and lens light.
In order to make more robust predictions for the LSST it is therefore necessary to include a more realistic simulation of the lens finding procedure, and possibly to revise our definition of a strong lensing event to take into account the positions of the multiple images with respect to the lens. 

\subsection{Overall properties} \label{sec:mock sample}

The procedure described above yielded a sample of $\Nlensfull$ lenses with $\theta_E>0\farcs 5$ for the entire MICE area of 5000 $\rm{deg}^2$, corresponding to a number density of 2.5 $\rm{deg}^{-2}$, comparable to the value of 2.1 $\rm{deg}^{-2}$ predicted by \citet{Son++23} with a similar strong lens event definition. 
Our number density is, however, about a factor of four smaller than that predicted by \citet{collett2015} for \Euclid. This difference is largely explained by our conservative definition of a strong lensing event, motivated by our choice of focusing on the most obvious lenses.
We found that the distribution of Einstein radii for the lens sample is approximately log-normal. It is therefore convenient to express the statistics in terms of $\tau_E\equiv \log(\theta_E/1'')$. For our baseline setup we obtained an average value of  $\mu_{\tau_E}^{\rm obs}=0.196$ and a dispersion 
$\sigma_{\tau_E}^{\rm obs}=0.239.$

\begin{figure*}
\centering
\includegraphics[width=1\hsize]{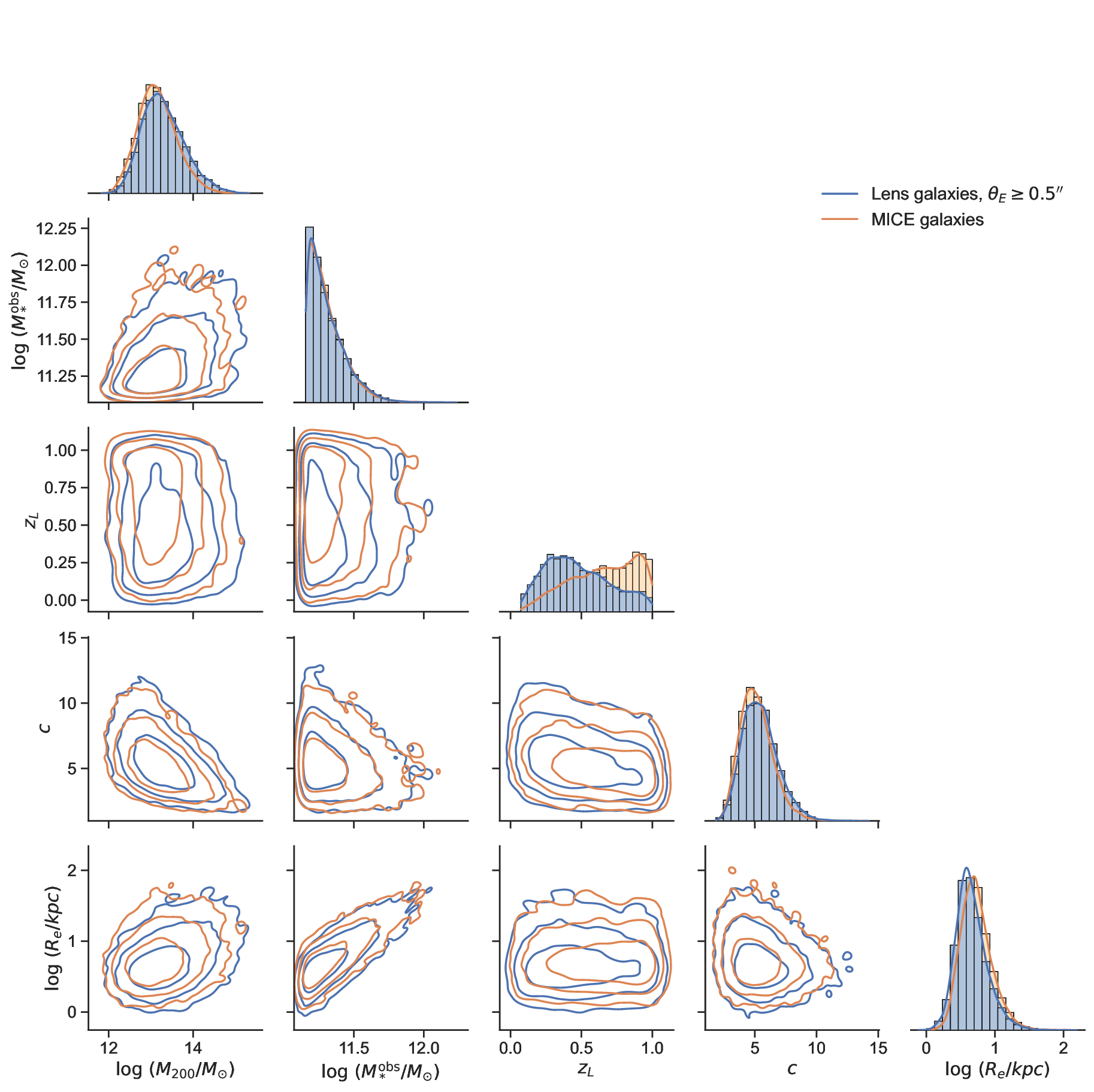}
\caption{The joint probability distributions of the mock observed sample (in blue) with $\theta_E \geq 0\farcs5$ and of the lens candidates selected from the MICE catalogues (in orange). The three contour lines delineate regions enclosing $68\%$ $95\%$ and $99.7\%$ of the distribution, respectively.}
\label{fig:diag1}
\end{figure*}

Fig. \ref{fig:diag1} shows the joint probability distributions of the lens properties of the observed \Euclid-like sample (in blue) with $\theta_E\geq 0\farcs5$, and the corresponding sample of the MICE galaxies before lensing (in orange). The results show that the lens sample is biased towards lower redshifts and larger halo masses compared to the MICE galaxies. This is expected, because the strong lensing cross-section area decreases with lens redshift but increases with halo mass \citep[see also][]{Son++23}. Moreover, the strong lensing selection prefers galaxies with larger stellar masses: higher probabilities are found in the tail of the stellar mass distribution of the lens galaxies compared to the MICE galaxies. This bias is much weaker than that in \cite{Son++23}, most likely the result of the difference in the stellar-to-halo mass relation.

\section{Results} \label{sec:results}
\subsection{Variations in the Einstein radius distributions}  \label{sec:bayes}

\begin{figure}
\centering
\includegraphics[width=1\hsize]{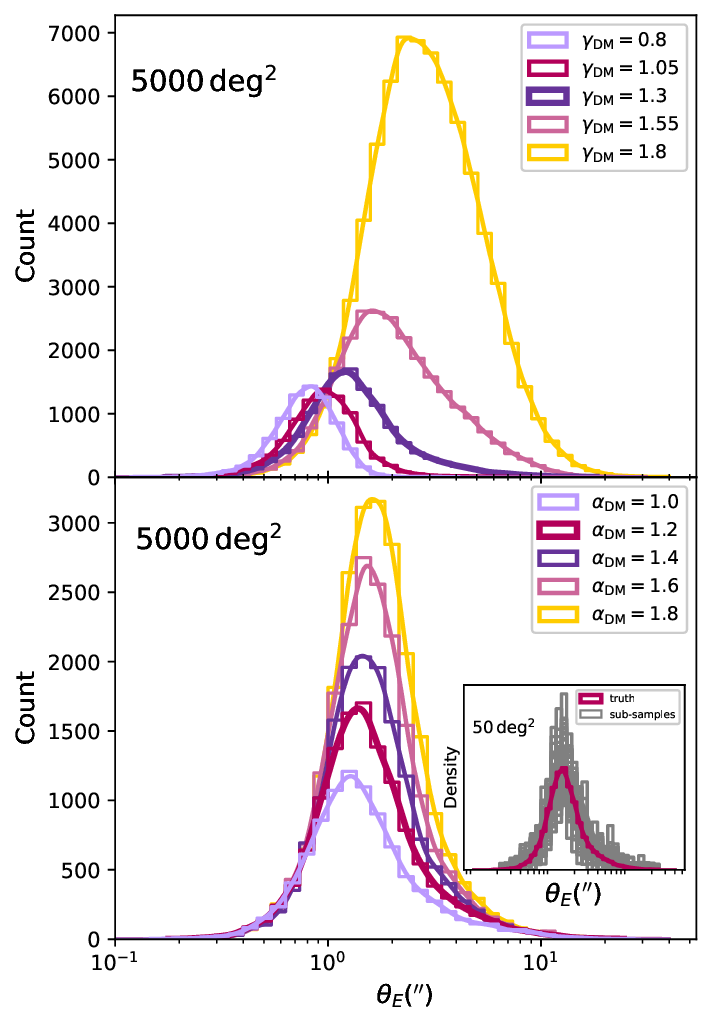}
\caption{The distributions of the Einstein radii for the observed mock sample of lenses with $\gamma_{\rm{DM}}^{\rm{true}}=1.3$ and $\alpha_{\rm{DM}}^{\rm{true}}=1.2$, and for realisations with the same $\alpha_{\rm{DM}}$ but varying $\gamma_{\rm{DM}}$ (top), the same $\gamma_{\rm{DM}}$ but varying $\alpha_{\rm{DM}}$ (bottom).}
\label{fig:einsrad}
\end{figure}

We fitted model populations of lenses to the distribution of Einstein radii of the mock observed sample. In order to fit observational data with simulations, we used the method described in \cref{sec:mock sample} to generate mock samples of strong lenses on an evenly-spaced $5\times 5$ grid covering the $\gamma_{\rm{DM}}$ - $\alpha_{\rm{SPS}}$ plane. 

The top panel of Fig. \ref{fig:einsrad} shows the distribution of the Einstein radius at $\alpha_{\rm{SPS}}=1.2$, for different values of $\gamma_{\rm{DM}}$. There is a clear correlation between $\gamma_{\rm{DM}}$ and the location, as well as the peak height of the distribution. When $\gamma_{\rm{DM}}$ increases from 0.8 to 1.8, the peak height increases by a factor $\sim 5$, while the median of the distribution moves from $0.8''$ to $3''$. Moreover, the distribution widens for larger $\gamma_{\rm{DM}}$. All in all, the moments of the distribution correlate with $\gamma_{\rm{DM}}$. The bottom panel shows the distribution of  Einstein radii for different values of $\alpha_{\rm{SPS}}$,
with $\gamma_{\rm{DM}}=1.3$. Compared to the top panel, the correlations between the moments of the Einstein radius distribution and $\alpha_{\rm{SPS}}$ are much weaker, except for the zeroth moment: the number of lenses increases by a factor $\sim 1.7$ when $\alpha_{\rm{SPS}}$ changes from 1 to $1.8$. 
This indicates that the additional information provided by the total number of lenses is especially helpful in constraining $\alpha_{\rm{SPS}}$.  

We divided the entire reference sample into 100 subsamples, each covering 1/100 of the entire MICE area, to
explore the constraining power of lens samples with a limited sample size, and to capture field-to-field variations. The inset plot of Fig. 2 illustrates the distributions of the Einstein radii for the full mock sample, as well as for the 100 subsamples. Notably, even with a few hundred lenses, the sample statistics of the truth can be replicated quite accurately. The average deviation from the mean of the subsample, compared to that of the whole sample, is approximately 1\%, while the deviation from the average of the subsample means to the mean of the full sample is 0.04\%. This close resemblance between the subsamples and the full sample enables us to effectively leverage a lens sample with a limited sample size to constrain the galaxy properties. 

Because we drew dark matter halos from the MICE catalogue when generating model lens populations, we implicitly assumed that we know the dark matter halo mass function and the mapping between halos and the observable properties of galaxies (in the case of MICE, the observed stellar mass). We also assumed that we know the properties of the background source population exactly. In a real analysis, these would be inferred by studying the distribution of unlensed galaxies in brightness and redshift space. 

Rather than fitting the full distribution in $\tein$, for simplicity we compressed its information content into the first three moments of the distribution of its logarithm, $\tau_{\rm E}$: the total number of lenses $\NSL$, the average, $\muE$, and the dispersion, $\sigE$.
The posterior probability distribution of the model parameters is then given by
\begin{equation}
\begin{aligned}
&\pr\left(\asps,\gammadm \mid\NSL,\muE,\sigE\right)  \propto \\ &\pr\left(\asps,\gammadm\right)\pr\left(\NSL,\muE,\sigE \mid\asps,\gammadm\right).
\end{aligned} \label{eq:posterior1}
\end{equation}

Assuming conditional independence given $\asps$ and $\gammadm$ between the lens observables $\NSL$, $\muE$, and $\sigE$, we have
\begin{equation}
\begin{aligned}
&\pr\left(\NSL,\muE,\sigE \mid\asps,\gammadm\right) \propto \\
&\pr\left(\NSL\mid\asps,\gammadm\right) 
\pr\left(\muE \mid\asps,\gammadm\right) 
\pr\left(\sigE \mid\asps,\gammadm\right). 
\end{aligned} \label{eq:posterior2}
\end{equation}

Therefore, by simulating mock samples for different values for $\asps$ and $\gammadm$, and expressing the lens observables $\NSL$, $\muE$, and $\sigE$ as functions of $\asps$ and $\gammadm$, we can obtain the three terms in equation \eqref{eq:posterior2} and consequently the posterior distribution in equation \eqref{eq:posterior1}.
For details on how we obtained the likelihood term we refer to Appendix \ref{app1}. 

\begin{figure*}
\centering
\includegraphics[width=1\hsize]{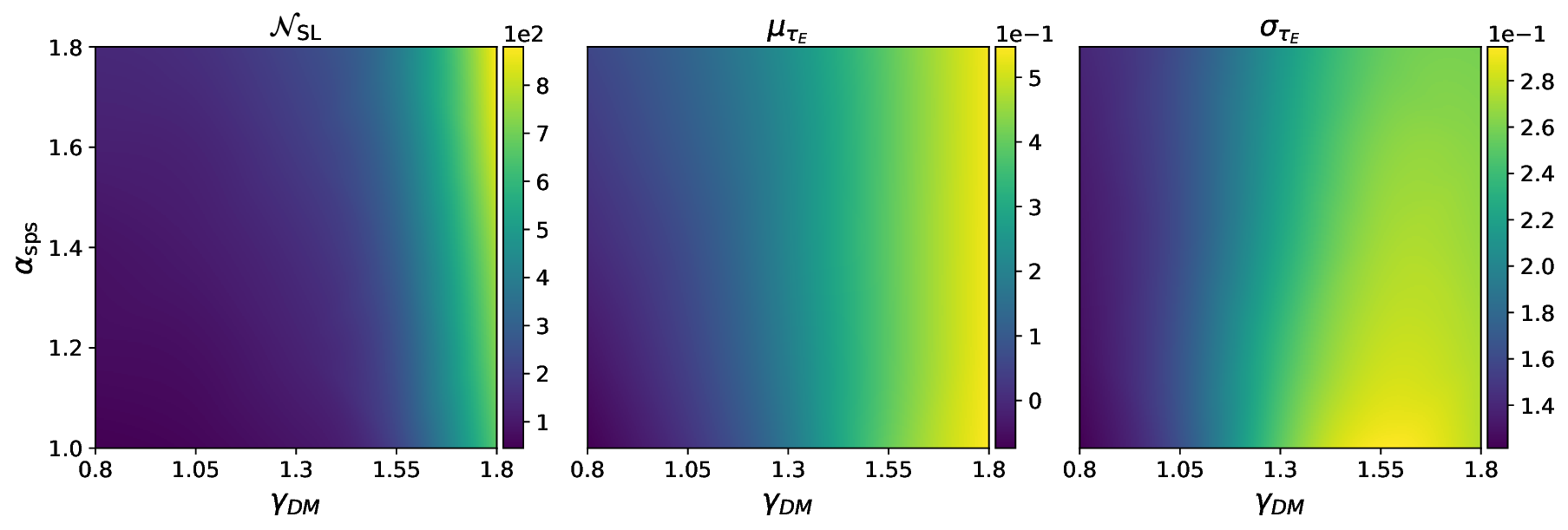}
\caption{The total number of strong lenses, the mean and the dispersion of the logarithm of the Einstein radius distributions of the mock samples, simulated on a $5\times 5$ grid (interpolated) on the $\gamma_{\rm{DM}}$ - $\alpha_{\rm{sps}}$ plane, as functions of $\alpha_{\rm{sps}}$ and $\gamma_{\rm{DM}}$.}
\label{fig:totnum} 
\end{figure*}

\subsection{Constraining galaxy properties}

Fig. \ref{fig:totnum} shows the first three moments (the total number of lenses $\NSL$, the mean $\muE$ and the dispersion $\sigE$) of the Einstein radius distributions of the mock samples as functions of both $\asps$ and $\gammadm$. There is a noticeable difference between the trend of $\NSL$ and that of the mean and the dispersion, especially at smaller $\gammadm$ and for $\asps \geq 1.3$. At larger $\gammadm$, the total number of lenses becomes less dependent on $\asps$. In the lower left corner at $\gammadm=0.8$ and $\asps=1$, the number density of lenses is 0.9 $\rm{deg}^{-2}$. In the upper right corner at $\gammadm=1.8$ and $\asps=1.8$, the lens number density becomes 17.6 $\rm{deg}^{-2}$. 

\begin{figure}
\centering
\includegraphics[width=1\hsize]{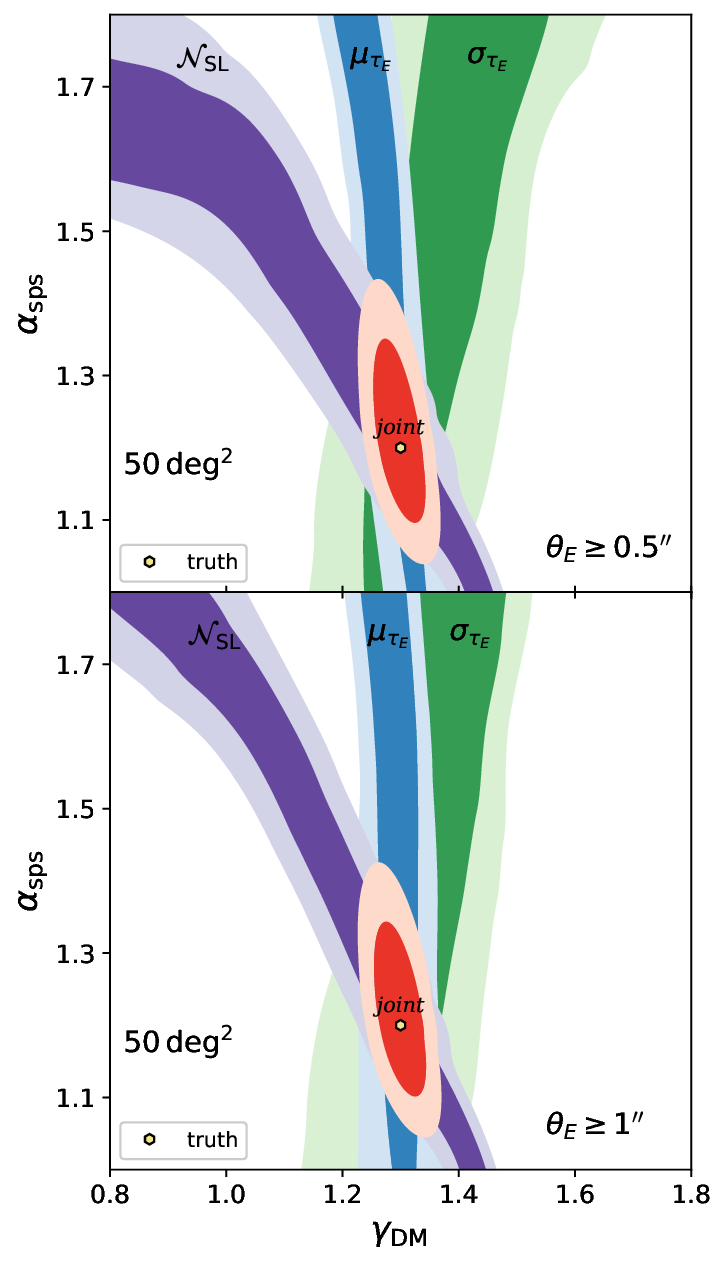}
\caption{The posterior distributions of the lenses with $\theta_E\geq 0.5^{\prime\prime}$ (top), and of $\theta_E\geq 1^{\prime\prime}$ (bottom), obtained as described in \cref{sec:bayes}. The two-level contour lines show respectively $68\%$ and $95\%$ probability masses, with the purple showing the constraint by $\NSL$, the blue by $\muE$ and the green by $\sigE$ .The red filled contours show the joint constraint of the other three. }
\label{fig:constrain_contour_cut}
\end{figure}

The top panel of Fig. \ref{fig:constrain_contour_cut} shows the posterior distributions $P\left(\asps,\gammadm\mid \NSL^{\rm{obs}}\right)$ (in purple), $P\left(\asps,\gammadm\mid \muE^{\rm{obs}}\right)$ (in blue), and $P\left(\asps,\gammadm\mid \sigE^{\rm{obs}}\right)$ (in green), of $\Nlens$ lenses with $\theta_E\geq 0.5^{\prime\prime}$ within a $50\,\rm{deg}^2$ area, which constitute our fiducial sample.
The estimated values and uncertainties ($68\%$ confidence) for $\alpha_{\rm{sps}}$ and $\gamma_{\rm{DM}}$ are $1.23^{+0.127}_{-0.127}$ and $1.30^{+0.045}_{-0.045}$, respectively.
These are remarkably small uncertainties: for reference, current estimates of $\asps$ in massive early-type galaxies can vary up to a factor of two, depending on the method used \citep{Newman++17}, and there is still no consensus on whether the inner dark matter slope in these galaxies is steeper or shallower than $\gammadm=1$ \citep{O+A18,Shajib2021,Tur++24}.
Our result highlights the power of working with a complete sample of lenses.

We also repeated the analysis by selecting only lenses with $\theta_E\geq 1^{\prime\prime}$, which corresponds to our more conservative scenario. 
The corresponding posterior probability distribution is shown in the bottom panel of Fig.~\ref{fig:constrain_contour_cut}. The results are very similar: this is a consequence of the fact that most of the simulated lenses have an Einstein radius larger than $1''$.
This is a promising result, as it means that high-precision measurements can still be obtained when focusing on the most obvious lenses, as long as the sample is complete.

\section{Sensitivity to cosmology}
\label{sec:cosmology}

Our analysis rests on the assumption that the cosmological parameters are known exactly, which is not the case in reality.
In this section we explore whether uncertainties in the cosmological parameters affect the estimation of the galaxy parameters that we wish to infer. We assume a flat geometry, because it is supported by observations and theoretical grounds, while any deviation will have a minimal impact. Moreover, our observables are insensitive to the Hubble parameter $H_0$. Therefore, we limit the discussion to the present-day mean matter density, $\Omega_{\rm m}$, and the normalisation of the linear matter power spectrum on a scale of $8h^{-1}$Mpc, $\sigma_8$.

One effect of varying $\Omega_{\rm m}$ is its impact on distance measurements and hence mass estimates. Over a narrow redshift range, a change in $\Omega_{\rm m}$ can be absorbed by a corresponding variation in $H_0$, which renders it not observable. Therefore only the redshift-dependent part of the sensitivity of the distance to $\Omega_{\rm m}$ matters. Given the current uncertainty from cosmic microwave background observations, distance ratios across the redshift range spanned by our simulation ($0.1 < z < 1.0$) vary by less than a percent with $\Omega_{\rm m}$. Such an uncertainty is much smaller than that on the stellar population synthesis mismatch parameter $\asps$ obtained with our sample of $\Nlens$ lenses and therefore can be neglected.

The main consequence of letting $\Omega_{\rm m}$ and $\sigma_8$ vary is the amount of clustering of dark matter on galaxy-scales, which in turn affects the number density of lenses \citep[see e.g.][]{boldrin2016}.
To quantify the amplitude of this effect we should repeat our analysis with different values of cosmological parameters and compare the results.
In practice this is difficult to do, because the dark matter halo masses in our models are assigned on the basis of the MICE simulations, which are only available for a fixed cosmology.
Instead, we use the following argument to estimate the impact of cosmology on the number of lenses.

\begin{figure*}
\centering
\includegraphics[width=0.9\hsize]{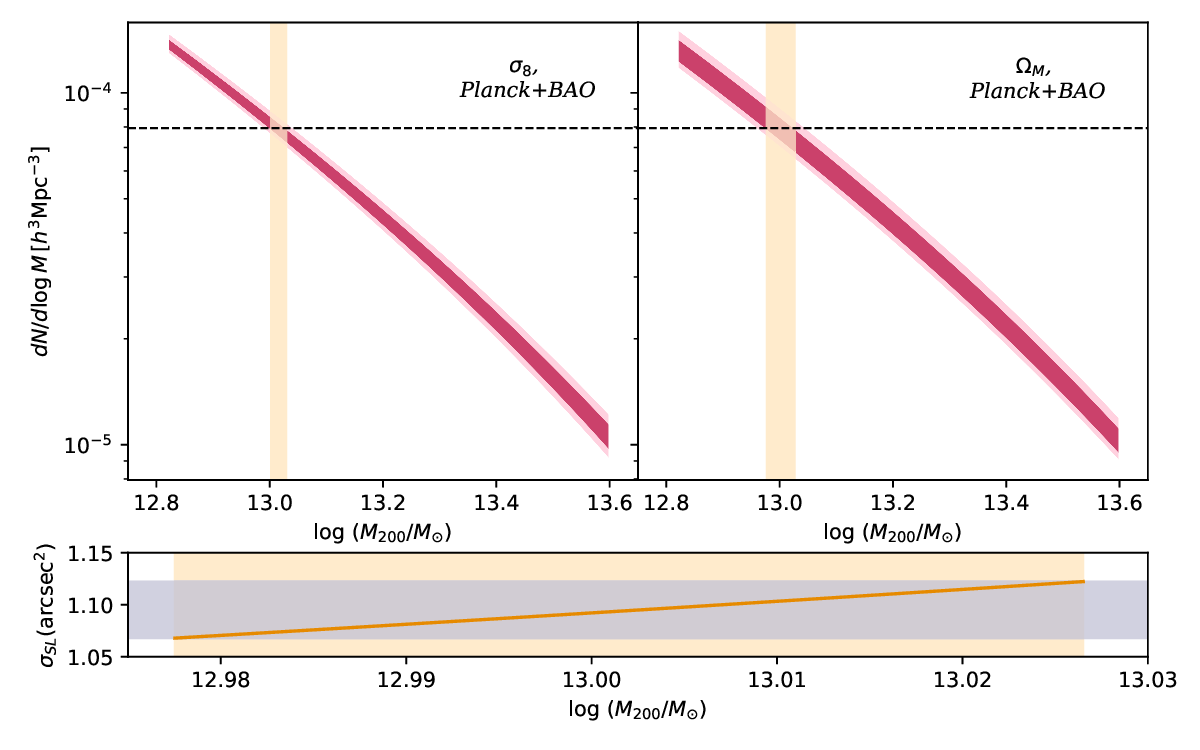}
\caption{Halo mass functions for the range in $\sigma_8$ (left) and $\Omega_{\rm m}$ (right), over the mass range $10^{12.8}-10^{13.6}\,\rm{M_{\odot}}$, as allowed by the cosmological parameter constraints from \cite{2020planck}. Halo mass functions were calculated using \texttt{Colossus} \citep{2018diemer}, assuming a mass function model of \cite{2010crocce}.
The bottom plot shows how the strong lensing cross-section area changes as a function of the halo mass in the relevant region, calculated from a lens with $M_h=10^{13}\,M_{\odot}$, $M_*=10^{11.5}\,M_{\odot}$, $\asps=$ 1.2, $\gammadm=$ 1.3, with a half-light radius of $R_{\rm{e}}=10$ kpc, at $z=1$. The test source is placed at $z=2.5$.}
\label{fig:cosmology}
\end{figure*}

The values for $\Omega_{\rm m}$ and $\sigma_8$ affect the number density of dark matter halos of a given mass. This can be seen in the top panel of Fig.~\ref{fig:cosmology}, which shows the variation in the halo mass function for the range in $\sigma_8$ and $\Omega_{\rm m}$ allowed by the uncertainties (68\% and 95\% confidence) by combining the {\it Planck} CMB constraints with those from BAOs \citep{2020planck}. We show the mass function for the mass range $10^{12.8}-10^{13.6}\,h^{-1}\,\rm{M_{\odot}}$ which is most relevant for the sample of strong lenses.
However, while the dark matter distribution changes, the stellar mass of the lenses does not, as it is fixed by the observations.
We then make the abundance matching ansatz: galaxies of a given stellar mass are assigned to halos with the same number density.
Then, a change in the number density of halos is equivalent to a shift in the value of the halo mass at fixed stellar mass.
As a result, the number of lenses does not vary as dramatically with cosmology as one might naively think by considering the effect on the halo mass function alone.

To predict the corresponding change in the number of lenses, we considered the lensing cross-section of a reference lens, with stellar mass $10^{11.5}\,M_{\odot}$, average half-light radius for its mass, and halo mass $10^{13}\,M_{\odot}$.
The bottom panel of Fig.~\ref{fig:cosmology} shows how the strong lensing cross-section changes over the relevant range in halo mass.
Varying $\Omega_{\rm m}$ and $\sigma_8$ within the $1\sigma$ ranges allowed by Planck measurements produces a change in the number density of halos that is equivalent to a shift of $\sim 0.05$ dex in halo mass. This, in turn, translates into a $\sim$ 2.4\% change in the lensing cross-section, which we take as an estimate of the variation in the number density of lenses. Such a variation is smaller than the Poisson fluctuation of our sample of $\Nlens$ lenses (in 50 deg$^2$), but would become comparable to it in a sample of 1736 lenses.
Hence, in a real-world application with a large sample, such as that expected by the end of the \Euclid mission, one would have to fit for parameters describing both galaxy structure and cosmology, thereby increasing the dimensionality of the problem. We leave the exploration of the dependence on cosmology to future work. 

\section{Discussion \& Conclusions} \label{sec:discussion}

In this work, we have demonstrated that it is possible to jointly constrain the dark matter inner slope $\gamma_{\rm{DM}}$ and the IMF mismatch parameter $\alpha_{\rm{sps}}$ using only the observed distribution of Einstein radii. 
One of the reasons why we are able to break the degeneracy between the two parameters is that we have implicitly assumed that we know the halo mass distribution of the foreground galaxy population.
Halo masses are important to predict the Einstein radius distribution, because, at fixed galaxy properties, both the Einstein radius and the strong lensing cross-section increase with increasing dark matter mass.

In our proof-of-concept, we assumed that the mapping between galaxy properties and halo mass is the same as that of the MICE simulation.
While it is unclear whether that model is sufficiently accurate for a real-world application, we can expect our understanding of the relation between galaxies and halos to improve in the near future, thanks to direct constraints on dark matter masses from upcoming weak gravitational lensing surveys.
However, some properties of the galaxy-halo connection, such as the intrinsic scatter, are notoriously difficult to constrain. 
Future work should examine the impact on the inferred parameters arising from possible inaccuracies in the model.
Nevertheless, the ability to use information on the halo mass distribution of the general galaxy population is a key feature of working with a complete sample of lenses. In contrast, when the selection function of a strong lens sample is not known, that information is difficult to use: this is because the halo masses of strong lenses are biased with respect to the general population, in a way that depends on the selection function \citep{Son++23}. 

We also assumed implicitly that the properties of the source population are known exactly and that they could be approximated as
point sources. As a consequence, all information depends on only two parameters: the redshift and the apparent magnitude. In reality, almost all galaxy sources are extended. Using the point source approximation means that we are ignoring the details of the surface brightness distribution. 
There are two factors that must be taken into consideration before applying this method to real lens samples. 
First, morphologies and the full surface brightness distribution of the sources must be incorporated in the lens modelling phase, to fully capture the complexity of real lenses. Second, the definition of a strong lens must be generalised in order to suit extended sources \citep[see section 2.4 of][for more information]{Son++23}.  \par

Our experiment rests on the assumption that it is possible to identify a complete and pure set of strong lenses, according to a well-defined lens definition criterion. In practice, this can be difficult to achieve: current samples of lens candidates are dominated by ambiguous systems with low confidence of being lenses \citep{2018sonnenfeld,Jacobs2019,2019petrillo,Jaelani2023}. One possible strategy to solve this problem is to focus on the most conspicuous and unmistakable lens candidates, such as lenses with well-resolved multiple images detected at a high signal-to-noise ratio, at the cost of a smaller sample size. 
Photometric redshifts could be used to improve the purity of the lens candidate pool, by flagging false-positives for which the supposed lensed images are at a redshift equal to or lower than that of the lens galaxy (such as spiral arms or tangentially elongated objects in the foreground).
However, blending between the lens light and background source makes the measurement of source colours challenging \citep{Lan++23}.
A better solution, but more demanding in terms of observational resources, would be to carry out a systematic spectroscopic follow-up campaign. Following up even a subset of candidates could be beneficial, as it would allow us to estimate the fraction of contaminants and correct the lens statistics accordingly.\par

In building the synthetic lens populations, we have made the simplifying assumption that all lens galaxies have the same $\alpha_{\rm{sps}}$ and $\gamma_{\rm{DM}}$. In reality this can be different: because the inner dark matter distribution is affected by baryonic physics, we expect a real sample of lenses to have a spectrum of dark matter inner slopes. The value of $\asps$ might also vary among galaxies, for example if the stellar IMF is not universal.
To first approximation, we expect that adding scatter to $\asps$ and $\gammadm$ would result in a larger value of the variance of the Einstein radius distribution. If not taken into account, this might bias the inference.
Assuming a wrong mass-concentration relation can also bias the inference. In particular, the concentration is degenerate with the inner slope $\gammadm$: given a gNFW profile, it is possible to increase the concentration and reduce $\gammadm$ while keeping the inner density profile, and hence the lensing properties, approximately unchanged. While this problem can be alleviated by incorporating weak lensing constraints \citep[see e.g.][]{2008mandelbaum}, it is difficult to completely eliminate the uncertainty on the concentration. Moreover, the relation between $\gammadm$, $\asps$ and the galaxy properties might be more complex: for example, they might scale with stellar mass, half-light radius or redshift.
In principle we could add free parameters to our model and apply the same inference method that we used in this paper to a real sample of lenses. In practice, however, this would require building a grid of models over a multi-dimensional space, which can be very challenging from a computational point of view. 
Another limitation of our approach is that, by fitting only the Einstein radius distribution, some valuable information is lost. For example, we expect the stellar mass to correlate with the Einstein radius, at fixed redshift and half-light radius, to a degree that depends on the value of $\asps$. Therefore it should be possible to make more precise and accurate inferences by taking into account such aspects of the data. 

In a real-world scenario, a better approach would be to fit a hierarchical model, trying to reproduce the observed Einstein radii on a lens-by-lens basis, instead of only fitting the Einstein radius distribution of the entire sample. \citet{2022sonnenfeld_paper3} showed how to fit such a model to a complete sample of lenses, in the context of power-law mass models. It should be possible to extend that formalism to allow for a more complex lens mass distribution, such as that used in this work.
The advantage of a hierarchical model approach is twofold. First, it allows for more flexibility in the parameterisation of the mass distribution of the galaxy population:  for instance, we could measure how the stellar IMF and dark matter distribution vary as a function of galaxy properties. Second, it makes use of all available information.
That approach would be particularly suitable to working with a much larger sample of lenses than the one considered in our simulation, such as that that will be obtained from the full \Euclid survey.
Nevertheless, our experiment has the merit of showing that the degeneracy between the stellar IMF and the dark matter inner slope can be broken with strong lensing data alone, which is a nontrivial result.
Although a similar conclusion was reached by \citet{2021sonnenfeld_paper1}, the main result of that work was based on the assumption that it is possible to measure the magnification ratio between two multiple images, for each lens. It is not clear how realistic that assumption is when applied to large datasets.
Our approach is more conservative, because Einstein radius measurements have been shown to be robust to a few percent \citep[see section 7.2 of][for a related discussion]{Eth++22}.
Moreover, the method of \citet{2021sonnenfeld_paper1} can only recover the properties of the lens galaxies, as opposed to those of the general galaxy population. The two are different due to strong lensing selection effects \citep[see][for an overview]{Son++23}. 

Despite the simplifications inherent in our lens modelling approach, our analysis suggests that the imperative lies not necessarily in increasing the sample size, but rather in having comprehensive knowledge of the selection. The fact that we have been able to effectively constrain the galaxy properties with a relatively modest sample size of a few hundred lenses is quite meaningful. This aspect holds particular significance for the forthcoming early data release of the Euclid deep fields, covering an area of approximately 50 deg$^2$ \citep{Mellier2024}. The ability to leverage our method in such a context underscores its potential utility in extracting valuable insights from real-world astronomical observations and data analysis, providing a valuable avenue for probing galaxy properties using strong lensing. 

In conclusion, our study shows that valuable information on key properties of the mass distribution of galaxies can be obtained from the Einstein radius distribution of a complete sample of lenses, even without any aid from stellar kinematics.
In order to extract that information, however, two key requirements must be met. First, a complete sample of lenses must be obtained: this poses new challenges upon lens finding methods. Second: the properties of the background source population must be known, in order to produce realistic simulations of lens samples to be fitted to the data.
We recommend upcoming strong lens surveys to address these challenges.

%
%
\bibliographystyle{aa}
\bibliography{ref}
\begin{appendix} 
\section{Bayesian inference model} \label{app1}
The likelihood functions describe the probabilities of observing $\NSL^{\rm{obs}}$, $\muE^{\rm{obs}}$ and $\sigE^{\rm{obs}}$, conditioned on $\asps$ and $\gammadm$. For any value of $\asps$ and $\gammadm$, these three functions provide the corresponding true number of lenses, the mean and the dispersion of the distribution in the logarithm of the Einstein radius. Hence, we can express the likelihoods as
\begin{equation}
\begin{aligned}
& P\left(\NSL^{\rm{obs}}\mid \asps,\gammadm\right) \propto P\left(\NSL^{\rm{obs}}\mid \NSL\left(\asps,\gammadm\right)\right)  \\
& P\left(\muE^{\rm{obs}}\mid \asps,\gammadm\right) \propto  P\left(\muE^{\rm{obs}}\mid\muE\left(\asps,\gammadm\right)\right) \\
& P\left(\sigE^{\rm{obs}} \mid \asps,\gammadm\right) \propto P\left(\sigE^{\rm{obs}} \mid \sigE\left(\asps,\gammadm\right)\right)
\end{aligned} \label{eq:likelihoods1}
\end{equation}
Since the likelihood function can be interpreted as the probability of obtaining $(\cdot)^{\rm{obs}}$ when drawing one random deviate from a known probability distribution $(\cdot)$, we make the following approximations of the likelihood functions as
\begin{equation}
\begin{aligned}
&P\left(\NSL^{\rm{obs}}\mid \NSL\left(\asps,\gammadm\right)\right) \approx \mathcal{P}\left(k=\NSL^{\rm{obs}}, \lambda=\NSL\right)\\
&P\left(\muE^{\rm{obs}}\mid \muE\left(\asps,\gammadm\right)\right) \approx  \mathcal{G}\left(x=\muE^{\rm{obs}},\mu=\langle\muE\rangle,\sigma=\sqrt{\rm{Var}[\muE]}\right) \\
&P\left(\sigE^{\rm{obs}} \mid \sigE\left(\asps,\gammadm\right)\right) \approx  \mathcal{G}\left(x=\sigE^{\rm{obs}},\mu=\langle\sigE\rangle,\sigma=\sqrt{\rm{Var}[\sigE]}\right)
\end{aligned} \label{eq:likelihoods2}
\end{equation}
where $\mathcal{P}(\cdot)$ denotes a Poisson distribution with rate $\lambda$, $\mathcal{G}(\cdot)$ denotes a Gaussian distribution with mean $\mu$ and standard deviation $\sigma$, and $\langle \cdot \rangle$ denotes the mean of a given quantity. So that now the likelihood function $P\left(\NSL^{\rm{obs}}\mid \NSL\left(\asps,\gammadm\right)\right)$ could be understood as the probability distribution of getting exactly $\NSL^{\rm{obs}}$, from a series of Poisson distributions with rate $\lambda=\NSL$. Similarly the other two likelihood functions are the probabilities of getting exactly $\muE^{\rm{obs}}$ and $\sigE^{\rm{obs}}$ from two Gaussian distributions. Using the approximate counterparts of the likelihood functions, we can implement Bayes theorem to jointly constrain $\asps$ and $\gammadm$.
\end{appendix}
\end{document}